 \documentclass[conference,romanappendices,letterpaper]{IEEEtran}
 \pdfoutput=1 
\usepackage{subfig}
\usepackage{stfloats}
\IEEEoverridecommandlockouts
\usepackage[ruled,linesnumbered]{algorithm2e}
\usepackage{algorithmic}
\SetKwInput{KwInput}{Input}                
\SetKwInput{KwOutput}{Output}              
\newcommand{\nonl}{\renewcommand{\nl}{\let\nl\oldnl}}
\usepackage{etoolbox}
\AtEndEnvironment{example}{\null\hfill$\square$}%
\AtEndEnvironment{remark}{\null\hfill$\square$}%

\usepackage{enumerate}
\usepackage{multirow,array}
\usepackage{cite}
\usepackage{graphicx}
\usepackage{psfrag}
\usepackage{array}
\usepackage{algorithm2e}
\usepackage{amssymb}
\usepackage{amsfonts}
\usepackage{float}
\usepackage{textcomp}
\usepackage{xcolor}
\usepackage{tabu}
\usepackage{array}
\usepackage{tabularx}
\usepackage{subfiles}

\usepackage{amssymb}
\usepackage{pifont}
\usepackage[amsthm,thmmarks]{ntheorem}

\newcolumntype{P}[1]{>{\centering\arraybackslash}p{#1}}
\newcolumntype{M}[1]{>{\centering\arraybackslash}m{#1}}

 \usepackage{amsmath}
\ExplSyntaxOn
\NewDocumentCommand{\RN}{m}
 {
  \textup{ \int_to_Roman:n { #1 } }
 }
\ExplSyntaxOff

\usepackage{float}

\newtheorem{conjecture*}{Conjecture}
\newtheorem{lemma}{Lemma}
\newtheorem{observation}{Observation}

\newtheorem{claim}{Claim}
\newtheorem{remark}{Remark}
\newtheorem{definition}{Definition}
\newtheorem{theorem}{Theorem}

\newcommand{\ff}{{\mathbb F}}
\newcommand{\unif}{{\mathcal U}}
\newcommand{\subfalph}{{\mathcal A}}
\newcommand{\sunjaf}{{\mathcal P}^{(M)}_{\mathsf{SJ}}}

\newcommand{\banuluk}{{\mathcal P}^{(M)}_{\mathsf{BU}}}
\newcommand{\sjmprime}{{\mathcal P}^{(M'+1)}_{\mathsf{SJ}}}

\newcommand{\bumprime}{{\mathcal P}^{(M'+1)}_{\mathsf{BU}}}
\newcommand{\wsj}{{\mathcal P}_{\mathsf{WSJ}}}
\newcommand{\wbu}{{\mathcal P}_{\mathsf{WBU}}}

\newcommand{\collsunjaf}{{\mathcal P}^{(M)}_{T-\mathsf{SJ}}}
\newcommand{\wcsj}{{\mathcal P}_{T-\mathsf{WSJ}}}
\newcommand{\expec}{{\mathbb E}}
\newcommand{\mileak}{\rho^{\mathsf{MIL}}}
\newcommand{\maxleak}{\rho^{\mathsf{MaxL}}}

\newcommand{\Prob}{\mathsf{P}}

\newcommand{\querysetsj}{{\cal Q}_\mathsf{SJ}}
\title{Sun-Jafar-Type Schemes for Weak Private Information Retrieval} 

\begin{document}


\author{Chandan Anand, Jayesh Seshadri, Prasad Krishnan, Gowtham R. Kurri 

}

\maketitle

\allowdisplaybreaks 

\begin{abstract}
In information-theoretic private information retrieval (PIR), a client wants to retrieve one desired file out of $M$ files, stored across $N$ servers, while keeping the index of the desired file private from each $T$-sized subset of servers. A PIR protocol must ideally maximize the rate, which is the ratio of the file size to the total quantum of the download from the servers, while ensuring such privacy. In Weak-PIR (WPIR), the criterion of perfect information-theoretic privacy is relaxed. This enables higher rates to be achieved, while some information about the desired file index leaks to the servers. This leakage is captured by various known privacy metrics. By leveraging the well-established capacity-achieving schemes of Sun and Jafar under non-colluding  ($T=1$) and colluding ($1<T\leq N$) scenarios, we present WPIR protocols for these scenarios. We also present a new WPIR scheme for the MDS scenario, by building upon the scheme by Banawan and Ulukus for this scenario. We present corresponding explicit rate-privacy trade-offs for these setups, under the mutual-information and the maximal leakage privacy metrics. In the collusion-free setup, our presented rate-privacy trade-off under maximal leakage matches that of the previous state of the art.  With respect to the MDS scenario under the maximal leakage metric, we compare with the non-explicit trade-off in the literature, and show that our scheme performs better for some numerical examples. For the $T$-collusion setup (under both privacy metrics) and for the MDS setup under the mutual information metric, our rate-privacy trade-offs are the first in the literature, to the best of our knowledge.
\end{abstract}

\let\thefootnote\relax\footnotetext{
Chandan, Jayesh, Dr. Krishnan and Dr. Kurri are with the Signal Processing and Communications Research Center, International Institute of Information Technology, Hyderabad, 500032, India (email: $\{$chandan.anand@research., jayesh.seshadri@research., prasad.krishnan@,gowtham.kurri@$\}$iiit.ac.in). 
}

\section{Introduction}
\label{sec:intro}




Privacy in communications has become a necessity in the present networked world. In Private Information Retrieval (PIR), a client seeks to retrieve a file from a library of files stored in one or more servers (or databases), in a manner that keeps the identity of the desired file private from each one (or a collection of subsets) of the servers. The nature of privacy could be information-theoretic or computational, the former being our focus here. A query-response protocol between the client and the servers that enables such private retrieval is called a \textit{PIR protocol}. A PIR protocol necessarily incurs an additional quantum of download from the servers apart from the desired file, due to the privacy requirement. The efficiency of a PIR protocol can be measured by its \textit{rate}, which is the ratio of the size of the desired file to the total quantum of downloaded bits. The \textit{capacity} of a given PIR system is the supremum of the rates overall PIR protocols. For a system with $N$ servers and $M$ files in which the servers do not communicate with each other regarding the queries (called the \textit{collusion-free} setup), the PIR capacity obtained by Sun and Jafar \cite{sun2017capacity} has the following expression. 
\begin{align}
\label{eqn:nocollcapacity}
    C_{\text{PIR}} = \left( 1 + \frac{1}{N} + \cdots + {\left(\frac{1}{N}\right)}^{M-1} \right)^{-1}.
\end{align}
The information-theoretically-private capacity-achieving scheme of Sun and Jafar uses the file size $L = N^M$, which was later improved \cite{sjARBITRARYmessagelengthNK-1} to $N^{M-1}$ via another protocol. Subsequently, Tian, Sun, and Chen  \cite{tian2019capacity} introduced a clever scheme (which we refer to as the \textit{TSC scheme}) for the collusion-free setup with $L=N-1$, which was shown to be optimal across a large class of codes. 

For the \textit{$T$-collusion} setting (in which any $T\leq N$ servers can collude, i.e. share queries), the PIR capacity was obtained \cite{sun2017capacityb} as follows, via an achievable scheme with $L=N^M$.
\begin{align}
\label{eqn:collcapacity}
    C_{T-\text{PIR}} = \left( 1 + \frac{T}{N} + \cdots + {\left(\frac{T}{N}\right)}^{M-1} \right)^{-1}.
\end{align}
Another setting studied in information-theoretic PIR consists of $N$ \textit{non-colluding} servers enabling distributed storage of coded versions of the $M$ files using an $(N,K)$ MDS code. The PIR capacity of such a system was obtained by Banawan and Ulukus \cite{banawan2018capacity}, accompanied by a scheme with $L=KN^M$ in the spirit of the Sun-Jafar scheme \cite{sun2017capacity}. This has the following expression. 
\begin{align}
\label{eqn:MDScapacity}
    C_{\text{MDS}} = \left( 1 + \frac{K}{N} + \cdots + {\left(\frac{K}{N}\right)}^{M-1} \right)^{-1}.
\end{align}
Taking ideas from the TSC scheme, PIR schemes for MDS-coded servers with much smaller file size were derived in \cite{Zhouetal_2019_TIT_MDS_optimalmsgsize_lowerupcost,Zhuetal_2020_MDS_optimalmessagesize_higher_upcost}, showing that $L=\mathsf{lcm}(K, N-K)$ is sufficient (and further, optimal, under certain conditions). Apart from these settings, a number of other scenarios and extensions have been studied in the context of PIR in recent years (see \cite{vithana_survey_PIR_appns_2023BITS}, for a nice survey). 

Information retrieval schemes which enable the client to successfully retrieve the desired file, without necessarily being information-theoretically private, are called \textit{weak-PIR} (WPIR) schemes. WPIR schemes enable the retrieval at higher rates than information-theoretically private retrieval protocols. WPIR schemes have been studied for the setup of collusion-free replicated servers in the works \cite{zhouetal_2020_WPIR_MaxL, lin2021multi, samy2021asymmetric} using a number of privacy metrics. These privacy metrics are defined based on various leakage measures between the queries at respective servers and the desired file index. These measures include \textit{mutual information leakage} \cite{lin2021multi}, \textit{maximal leakage} \cite{issa2019operational}, \textit{worst-case information leakage} \cite{kopf2007information}, \textit{$\epsilon$-privacy} \cite{samy2021asymmetric}, and the \textit{converse-induced privacy metric} \cite{Chenetal_2024ISIT_CIPM_WPIR}. 
By enhancing the schemes in \cite{zhouetal_2020_WPIR_MaxL, lin2021multi}, WPIR schemes with improved rate-privacy trade-offs under the mutual information and maximal leakage metrics were recently obtained \cite{improved2022wpir}; these schemes achieve the best-known rate-privacy trade-offs for the collusion-free setting with replicated servers, to the best of our knowledge. The schemes in \cite{improved2022wpir} involve time-sharing a `clean' download phase with a variation of the TSC scheme. In the clean download phase,  the client directly downloads the desired file from a randomly chosen server, thus leaking the index completely to only that server.  The scheme in \cite{improved2022wpir} was also recently extended to handle heterogenously trusted servers \cite{Huangetal_ISIT2024_WPIR_HetTrustServers}. Based on the TSC scheme, WPIR schemes were also proposed recently \cite{orvedal2024weaklyprivateinformationretrievalmdscoded} for MDS-coded servers. There also exists a considerable body of literature involving single-server WPIR (for instance, \cite{hylin_2021_singleservercapacityWPIR,Yakimenka_ISIT2021_SingleServer_RateDistortion}). 

The contributions and organization of this work are as follows. In Section \ref{sec:sjwpircollusionfree}, we propose WPIR schemes for the collusion-free setup with replicated servers, and obtain the explicit rate-privacy trade-offs achieved by these schemes under the mutual information and maximal leakage metrics (Theorems \ref{thm:WSJ_MIL} and \ref{thm:WSJ_MaxL}). The proposed WPIR schemes are based on the simple idea of time-sharing the Sun-Jafar scheme \cite{sun2017capacity} with a clean download phase. Surprisingly, this simple idea recovers the best rate-privacy trade-off known so far \cite{improved2022wpir} under the maximum leakage metric, for the collusion-free replicated server setup. For the mutual information metric, we obtain a trade-off which is suboptimal when compared to the best known \cite{improved2022wpir}. 

Subsequently, in Section \ref{subsec:MDScodedWPIR}, we present WPIR schemes for the collusion-free, MDS-coded setup, using the time-sharing idea in conjunction with the scheme by Banawan and Ulukus \cite{banawan2018capacity}. We characterize the corresponding explicit rate-privacy trade-offs in Theorems \ref{thm:WBU_MIL} and \ref{thm:WBU_MaxL}. To the best of our knowledge, the only known WPIR schemes for MDS-coded servers are from \cite{orvedal2024weaklyprivateinformationretrievalmdscoded}, which considers only the maximal leakage metric and further does not provide an explicit expression for the rate-privacy trade-offs achieved. For some examples provided in \cite{orvedal2024weaklyprivateinformationretrievalmdscoded}, we observe through numerical calculations that our trade-off performs better. 

Finally, in Sub-section \ref{subsec:Tcoll-WPIR}, by harnessing the time-sharing idea together with the Sun-Jafar $T$-collusion scheme \cite{sun2017capacityb}, we present WPIR schemes that handle $T$-collusion and characterize the rate-privacy trade-offs these achieve in Theorems \ref{thm:TWSJ_MIL} and \ref{thm:TWSJ_MaxL}. To the best of our knowledge, this is the first proposed WPIR scheme for $T$-collusion.

\textit{Notation:} For $a,b\in {\mathbb Z}$ such that $a\leq b$ we denote the set of integers $\{a,a+1,\hdots, b\}$ as $[a:b]$. For any $x \in {\mathbb R}$ we denote the positive value of $x$ by $(x)_{+} \triangleq \max(0, x)$. The notation $\ff$ denotes a finite field. $\unif(A)$ denotes the uniform distribution on the set $A$. The expectation of the random variable $X$ is denoted by $\expec[X]$. For a set $A$, the set of all $b$-sized subsets of $A$ is denoted by $\binom{A}{b}$. Let $\cal S$ denote a set indexing items $X_i:i\in \cal S$. For a subset $\cal T\subseteq \cal S$,  we denote the subset of items $X_i:i\in \cal T$ as $X_{\cal T}$.
Random quantities are denoted using capitals; whereas realizations of random variables are denoted in small letters. Vector quantities (specifically, realizations of random vectors) are in bold. All logarithms are taken base $2$ unless otherwise specified.
\section{PIR System Model and Preliminaries}
\label{sec:systemmodelandprelim}

An information retrieval system consists of a client and $N$  servers (or databases), which are indexed throughout this work as $[0:N-1]$. In the replicated-server setup, each server stores all $M$ files, denoted by $W_1,W_2,\dots,W_M$, which are independent and identically distributed.  The client wishes to retrieve one of the $M$ files, denoted by $W_{\theta}$, where the index $\theta$ of the desired file is assumed to be distributed as $\theta \sim \unif([1:M])$. We represent the files as $W_i = (W_{i,1}, W_{i,2}, \cdots, W_{i,L}), i\in [1:M]$, where we assume $W_{i,j}, \forall i,j,$ are independent and identically distributed according to $\unif(\subfalph)$, where $\subfalph$ is an abelian group. Thus, we have their joint entropy $H(W_{[1:M]})=H(W_1, W_2, \cdots, W_M)=\sum^M_{i=1}H(W_i)=ML\log|\subfalph|$.


To retrieve the desired file $W_\theta$, the client initiates an \textit{information retrieval} protocol (or an \textit{IR scheme}). In such an IR protocol, 
the client generates queries $Q_l^{\theta}: l\in[0:N-1]$ where the query $Q_l^{\theta}$ is chosen from a set $\mathcal{Q}_l$, as a function of the file index $\theta$ and random bits privately available at the client. The query $Q_l^{\theta}$ is then sent to server $l$. The server $l$ then responds with an answer string $A_l^{\theta}$, which is a deterministic function of the query $Q_l^{\theta}$ and the data it stores. We denote the length of $A_l^{\theta}$ as  $\ell(A_l^{\theta})$ bits. 
An IR protocol should be \textit{correct}, i.e., enable the client to decode the desired file from the queries and answers. Thus, 
\begin{align}
\label{eqn:correctness}
    H(W_{\theta} | A^{\theta}_{[0:N-1]}, Q^{\theta}_{[0:N-1]}) = 0.
\end{align}
The rate of an IR protocol is defined as follows.
 \begin{align}
 \label{eqn:rate}
     R  \triangleq \frac{L\log|\subfalph|}{\expec[D]}, 
 \end{align}
where $D=\sum_{l=1}^N \ell(A_l^{\theta})$ 
is the total downloaded bits and the expectation is over the distribution of the queries $Q_{[0:N-1]}^{\theta}$ and $\theta$. For $T\in[1:N]$, an IR protocol is said to be a {$T$-collusion private information retrieval} protocol (or simply, a \textit{$T$-PIR scheme}), if the queries are designed in such a way that they do not disclose the index $\theta$ to any subset of $T$ servers. In particular, the following \textit{privacy constraint} must be satisfied for every ${\cal T}\subseteq [0:N-1]$ such that $|{\cal T}|=T$: 
\begin{align}
\label{eqn:Tcollusionprivacy}
I(Q^{\theta}_{\mathcal{T}}, A^{\theta}_{\mathcal{T}}, W_{[1:M]}; \theta) = 0.
\end{align}
When $T=1$, the PIR setup is called a \textit{collusion-free} setup. 

An information retrieval scheme that does not necessarily satisfy \eqref{eqn:Tcollusionprivacy} but satisfies \eqref{eqn:correctness} is called a Weak-PIR (WPIR) scheme. In this context, those schemes which satisfy both \eqref{eqn:correctness} and \eqref{eqn:Tcollusionprivacy} are said to satisfy \textit{perfect privacy}. In this work, we consider the privacy metrics for WPIR based on mutual information and maximal leakage measures similar to \cite{zhouetal_2020_WPIR_MaxL,lin2021multi,improved2022wpir}. The mutual information leakage at server $l$ is $I(Q_l^{\theta};\theta)$, while the maximal leakage at server $l$ is $\mathsf{MaxL}(\theta,Q_l^{\theta})\triangleq \log\left(\sum_{\pmb{q}\in{\mathcal Q}_l}\max_{m\in[1:M]}\Prob_{Q_l^{\theta}|\theta}(\pmb{q}|m)\right).$ Based on these, the corresponding privacy metrics in WPIR are defined as follows. 
\begin{definition}\cite{zhouetal_2020_WPIR_MaxL,lin2021multi}
\label{defn:privacymetrics_WPIR_nocollusion}
The mutual information based privacy metric for a given IR scheme $\cal C$ with queries $Q_l:l\in[0:N-1]$ is defined as follows. 
\begin{align}
\label{eqn:MI_leakage_definition}
     \mileak({\cal C}) \triangleq \frac{1}{N}\sum_{l=0}^{N-1} I(\theta;Q_l^{\theta}).
 \end{align}
The maximal leakage based privacy metric of IR scheme $\cal C$ is defined as follows.
\begin{align}
\label{eqn:defn:maxlprivacy}
\maxleak({\cal C})&\triangleq\max_{l\in[0:N-1]}\log\left(\sum_{\pmb{q}\in{\mathcal Q}_l}\max_{m\in[1:M]}\Prob_{Q_l^{\theta}|\theta}(\pmb{q}|m)\right).
\end{align}
\end{definition}

\subsection{Summary of Sun-Jafar \cite{sun2017capacity} and TSC \cite{tian2019capacity} schemes}
Sun and Jafar \cite{sun2017capacity} presented the breakthrough result for collusion-free PIR with $N$ servers and $M$ files, demonstrating a scheme that achieves the capacity \eqref{eqn:nocollcapacity} and a matching converse. To execute the PIR protocol in \cite{sun2017capacity}, which we will denote as $\sunjaf$, we need the file size $L=N^M$. The achievability scheme presented in \cite{sun2017capacity} queries for $(\sum_{s=1}^M N^s)$ linear combinations of the $N^M$ segments of the files systematically, while ensuring conditions \eqref{eqn:correctness} and \eqref{eqn:Tcollusionprivacy}. 
From the description of the protocol $\sunjaf$ in \cite{sun2017capacity}, and due to the fact that it is private, we have the following observation.
\begin{observation}
\label{obs:sunjafarquerydistr}
Let $\querysetsj^{[1:M]}$ denote the set ${\cal Q}_0$ of all possible queries as seen by the server with index $0$, in the protocol $\sunjaf$ executed on the files with indices $[1:M]$. The following are then true. 
\begin{itemize}
\item Each $\pmb{q}\in\querysetsj^{[1:M]}$ denotes non-zero linear combinations involving segments of all $M$ files in the system.
\item ${\cal Q}_l=\querysetsj^{[1:M]}, \forall l\in[0:N-1]$.
\item For all $\pmb{q}\in\querysetsj^{[1:M]}$ and $m\in[1:M]$, the probability $\\Prob_{Q_l^{\theta}|\theta}(\pmb{q}|m)=c_M>0$, where $c_M$ is a constant that depends only parameters $M$ and $N$  (see \cite[Lemma 3]{sun2017capacity}, for a proof). 
\item Further, $H(\theta|Q^{\theta}_l)=\log M$. 
\end{itemize}
\end{observation}
 Similar properties can be observed for the $T$-collusion PIR scheme presented in \cite{sun2017capacityb} and the MDS-PIR scheme in \cite{banawan2018capacity}. We use these observations to prove some claims in the present work.
 

In $\sunjaf$, the total number of symbols downloaded from any particular server is the same across all possible queries, however this is not the case in the TSC scheme \cite{tian2019capacity}. This subtle difference is exploited in the TSC scheme \cite{tian2019capacity} to obtain the exponential reduction in the file size ($L=N-1$, in this scheme), while achieving the PIR capacity \eqref{eqn:nocollcapacity}. The TSC scheme also has the minimum upload cost among all schemes in a large class. The essential idea of the TSC scheme is that the client requests a random linear combination of the segments of all files from one server, and uses it as a `random mask' for the download of desired segments from other $N-1$ servers. The probability distribution of this random mask linear combination is carefully chosen to ensure that the capacity of PIR is achieved. MDS-PIR schemes based on similar ideas were also obtained subsequently \cite{Zhouetal_2019_TIT_MDS_optimalmsgsize_lowerupcost,Zhuetal_2020_MDS_optimalmessagesize_higher_upcost}. Further, almost all known WPIR schemes \cite{zhouetal_2020_WPIR_MaxL,improved2022wpir,lin2021multi,Huangetal_ISIT2024_WPIR_HetTrustServers, Chenetal_2024ISIT_CIPM_WPIR,orvedal2024weaklyprivateinformationretrievalmdscoded} in literature were essentially derived from the TSC scheme with different probability assignments for the random mask linear combination and time-sharing it with a `clean' download phase, in which the complete file is downloaded from a single randomly chosen server. 
\subsection{Privacy Metrics for PIR under Server Collusion}
In this work, we also propose a WPIR scheme for server collusion scenarios with replicated storage. We formally provide the privacy metrics for this scenario, as they seem unavailable in literature so far. These are natural extensions of \eqref{eqn:MI_leakage_definition} and \eqref{eqn:defn:maxlprivacy} in Definition \ref{defn:privacymetrics_WPIR_nocollusion}. For a given $T$-collusion WPIR protocol $\cal C$, we have the following definitions for the mutual information and maximal leakage privacy metrics.
\begin{align}
\label{eqn:MI_leakage_definition_Tcoll}
     \mileak({\cal C}) \triangleq \frac{1}{\binom{N}{T}}\sum_{{\cal T}\in\binom{[0:N-1]}{T}} I(\theta;Q_{\cal T}^{\theta}),
 \end{align}
 where $I(\theta;Q_{\cal T}^{\theta})=I(\theta;\{Q_t^{\theta}:t\in {\cal T}\})$.
\begin{align}
\label{eqn:defn:maxlprivacy_Tcoll}
\maxleak({\cal C})&\triangleq\max_{{\cal T}\in\binom{[0:N-1]}{T}}\log\left(\sum_{\pmb{q}_{\cal T}\in{\mathcal Q}_{\cal T}}\max_{m\in[1:M]}\Prob_{Q_{\cal T}^{\theta}|\theta}(\pmb{q}_{\cal T}|m)\right).
\end{align}

\section{Sun-Jafar WPIR Scheme for Collusion-Free Scenario}
\label{sec:sjwpircollusionfree}


We now present a natural extension of $\sunjaf$ to WPIR. The proposed protocol works essentially by time-sharing a `clean' download protocol as was considered in \cite{improved2022wpir}  and the Sun-Jafar protocol on a randomly selected subset of non-desired files along with the desired file. We present the trade-offs these achieve and compare them with the existing best known ones \cite{improved2022wpir} in literature. 


\subsection{WPIR scheme based on $\sunjaf$}
\label{subsec:WPIRSJ_queryserverresponses}
We denote our WPIR protocol for the collusion-free setup as $\wsj$. We assume, as in $\sunjaf$, each file is divided into $L=N^M$ segments. For each $s\in[2:M]$, we assume a partitioning of the collection of segments of each file into $N^{s}$ super-segments, with each super-segment comprising $N^{M-s}$ original segments taken in order. Of course, the super-segments are exactly the segments when $s=M$. These super-segments are naturally indexed as $[1:N^s]$. For convenience, we shall refer to this as \textit{$s$-level super-segmentation.} We now describe the query-generation of $\wsj$.
\subsubsection{Query Generation of $\wsj$}
Assume the desired file index is $\theta=k$. The client picks the value of a random variable $M'$ according to some chosen distribution $\Prob_{M'}$ on the set $[0:M-1]$. The value of $M'$ will denote the number of additional non-desired files chosen for the execution of the protocol. The query generation in the $\wsj$ protocol proceeds as per the two cases. 
\begin{itemize}
    \item \textit{Case 1:} If $M'=0$, the client selects a server index $S$ according to $\unif([0:N-1])$. The client sends query $\#_k$ to server-$S$ and sends query $\phi$ (null query) to all other servers.
    \item \textit{Case 2:} If $M'=m'\neq 0$, the client selects $m'$ distinct files uniformly at random from $\{W_j : j\in[1:M]\backslash \{k\}\}$ denoted as $\{W_j, j\in J\}$, where $|J|=m'$. The Sun-Jafar ${\cal P}_\mathsf{SJ}^{m'+1}$ query-generation protocol is then executed for $N$ servers, on the set of files $\{W_j:j\in J\}\cup \{W_k\}$, the desired file index $k$, and using $(m'+1)$-level super-segmentation.
\end{itemize}

\subsubsection{Server Responses} As per the query generation protocol, if $M'\neq 0$, then any query $\pmb{q}$ seen by any server $l\in[0:N-1]$ lies in the set $\querysetsj^{(J\cup\{k\})}$, as denoted in Observation \ref{obs:sunjafarquerydistr}. 
For any such query $\pmb{q}$, let ${\cal M}_{\pmb{q}}$ denote the set of file indices that appear in $\pmb{q}$. Observe that ${\cal M}_{\pmb q}$ is known to the server which sees $\pmb{q}$. Specifically, ${\cal M}_{\pmb{q}}=J\cup\{k\}$, if and only if $\pmb{q}\in\querysetsj^{(J\cup\{k\})}$; in this case, $|{\cal M}_{\pmb{q}}|=|J|+1=M'+1$.
The server-response protocol at server $l$ for $\wsj$ is described as follows.
\begin{itemize}
    \item If $Q_l = \#_k$, server $l$ sends the entire $W_k$ to the client.
    \item If $Q_l = \phi$, server $l$ sends no response.
    \item For any $\pmb{q}\in \querysetsj^{(J\cup\{k\})}$ (which occurs when $M'\neq 0$), the server responds according as per the $\sjmprime$ protocol, on files with indices ${\cal M}_{\pmb{q}}=J\cup\{k\}$, and $|{\cal M}_{\pmb{q}}|$-level super-segmentation. 
\end{itemize}

\subsection{Correctness and Rate} The correctness condition is satisfied when $M'=0$, as in this case the server receiving the query $\#_k$ directly responds with the desired file $W_k$. Otherwise, when $M'\neq 0$, the server provides a response as per the Sun-Jafar protocol \cite{sun2017capacity}, with $M'+1$ files. Therefore, the correctness of the $\wsj$ protocol follows directly from the arguments proved in \cite{sun2017capacity}.

Now to calculate the rate of $\wsj$. When $M'=0$, the download cost is clearly $N^M$. When $M'=m'\neq 0$, the Sun-Jafar protocol ${\cal P}_{\mathsf{SJ}}^{(m'+1)}$  is utilized (with $(m'+1)$-level super-segmentation, with super-segments of size $N^{M-(m'+1)}$. Thus, the download cost in this case $(\sum_{s=1}^{m'+1}N^s)(N^{M-(m'+1)}) = \sum_{s=M-m'}^M N^s$, following arguments in \cite{sun2017capacity}.
Thus, the rate of $\wsj$ can be derived using \eqref{eqn:rate} as follows:
\begin{align}
    R_{\mathsf{WSJ}} 
    &\overset{}{=} \frac{N^M}{\expec \left[\sum_{s=M-M'}^M N^s\right]}\overset{}{=} \frac{1}{\expec\left[(1 + N^{-1} + \cdots + N^{-M'})\right]}\nonumber\\
    &\overset{}{=} \frac{1-\frac{1}{N}}{\expec\left[(1 - \frac{1}{N^{M'+1}})\right]}
\label{eqn:rategeneralpm'}
    \overset{}{=} 
    \frac{1-\frac{1}{N}}{1-\expec\left[{\frac{1}{N^{M'+1}}}\right]},
\end{align}
where the expectation is with respect to the distribution of $M'$.

\subsection{Privacy metrics for arbitrary $\Prob_{M'}$}
We now derive the privacy metrics $\maxleak$ and $\mileak$ for the $\wsj$ protocol. To do this, we need the following claim.
\begin{claim}
    \label{claim:thetagivenQ}
    With respect to the protocol $\wsj$, the following statements are true for any $m\in[1:M]$. 
    \begin{enumerate}
        \item For any $k\in[1:M]$, $$\Prob_{\theta|Q_l^{\theta}}(m|\#_k)=\begin{cases}1&\text{if}~m=k\\
        0&\text{if}~m\neq k.\end{cases}$$
        \item $\Prob_{\theta|Q_l^{\theta}}(m|\phi)=\frac{1}{M}$. 
        \item For $\pmb{q}\notin \{\phi\}\cup\{\#_k:k\in[1:M]\}$,
    $$\Prob_{\theta|Q_l^{\theta}}(m|\pmb{q})=\begin{cases}\frac{1}{|{\cal M}_{\pmb{q}}|}, &\text{if}~m\in {\cal M}_{\pmb{q}},\\
        0&\text{if}~m\notin {\cal M}_{\pmb{q}}.\end{cases}$$
    \end{enumerate}
\end{claim}
\begin{IEEEproof}
The proof for parts \textit{(1)} and \textit{(2)} are straightforward. Hence we show only the part \textit{(3)}. 

Note that the desired file index $\theta$ must be in ${\cal M}_{\pmb q}$, if 
$Q_l^{\theta}=\pmb{q}$. Specifically, if $\pmb{q}\notin \{\phi\}\cup\{\#_k:k\in[1:M]\}$, then $\pmb{q}\in\querysetsj^{J\cup\{k\}}$ for desired file index $k$ and for $J\subset [1:M]\setminus\{k\}$ such that $J\cup\{k\}={\cal M}_{\pmb q}$.  Thus, if $m\notin {\cal M}_{\pmb{q}}$, then $\Prob_{Q_l^\theta|\theta}(\pmb{q}|m)=0$, which implies $\Prob_{\theta|Q_l}(m|\pmb{q})=0$. 

Now, if $m\in {\cal M}_{\pmb{q}}$, by Observation \ref{obs:sunjafarquerydistr}, we have $\Prob_{Q_l^\theta|\theta}(\pmb{q}|m)=c_{|{\cal M}_{\pmb{q}}|},$ where $c_{|{\cal M}_{\pmb{q}}|}$ is a constant that depends only on $|{\cal M}_{\pmb{q}}|$ and $N$. Now, as $\Prob_\theta$ is uniform, we have $\Prob_{\theta|Q_l^\theta}(m|\pmb{q})=\frac{\Prob_{Q_l^\theta|\theta}(\pmb{q}|m)}{\sum_{m_1\in[1:M]}\Prob_{Q_l^\theta|\theta}(\pmb{q}|m_1)}=\frac{c_{|{\cal M}_{\pmb{q}}|}}{|{\cal M}_{\pmb{q}}|c_{|{\cal M}_{\pmb{q}}|}}=\frac{1}{|{\cal M}_{\pmb{q}}|}$. This completes the proof.
\end{IEEEproof}

Using Claim \ref{claim:thetagivenQ}, we now give expressions for the privacy metrics in the following lemmas.
\begin{lemma}
\label{lemma:WSJ_MI_Leakage}
     \begin{align}
     \label{eqn:MIleak_generalM'_WSJ}
        \mileak(\wsj)= &\left(1 - \bigg(1-\frac{1}{N}\bigg)\Prob_{M'}(0)\right) \log M\nonumber\\& - \expec[\log(M'+1)].
    \end{align}
\end{lemma}
\begin{IEEEproof}
We show that $I(\theta; Q_{l}^{\theta})$ is equal to the R. H. S of \eqref{eqn:MIleak_generalM'_WSJ}, for all $l$. Using \eqref{eqn:MI_leakage_definition}, the statement follows. 
    \begin{align}
        I(\theta; &Q_{l}^{\theta}) \\
        \overset{}{=}& H(\theta) - H(\theta | Q_{l}^{\theta})\nonumber\\ 
        \overset{(a)}{=}& \log M - \sum_{\pmb{q} \in \mathcal{Q}_{l}} \Prob_{Q_{l}^{\theta}}(\pmb{q})H(\theta | Q_{l}^{\theta}=\pmb{q})\nonumber\\
        \label{eqn:mileakcalculation_intermed_sj}
        \overset{}{=}& \log M - \biggl(\sum_{m\in {[1:M]}} \Prob_{Q_{l}^{\theta}}(\#_m)H(\theta|Q_{l}^{\theta} = \#_m)\nonumber\\ &~~~~~~~~~+ \Prob_{Q_{l}^{\theta}}(\phi)H(\theta|Q_{l}^{\theta} = \phi)\nonumber\\ &~~~~~~~~~+ \sum_{m' = 1}^{M-1} \sum_{\substack{\pmb{q}\in \mathcal{Q}_{l}: \\|{\cal M}_{\pmb{q}}|=m'+1}}\Prob_{Q_{l}^{\theta}}(\pmb{q})H(\theta|Q_{l}^{\theta}=\pmb{q}) \biggl).
        \end{align}
        Here $(a)$ holds as $\theta$ is chosen uniformly at random from $[1:M]$. Now, using Claim \ref{claim:thetagivenQ}, we have $H(\theta|Q_{l}^{\theta} = \#_m)=0$, $H(\theta|Q_{l}^{\theta} = \phi)=\log M$, and if $|{\cal M}_{\pmb{q}}|>1$, $H(\theta|Q_{l}^{\theta}=\pmb{q})=\log |{\cal M}_{\pmb{q}}|$. Further, $\Prob_{Q_{l}^{\theta}}(\phi)=\Prob_{M'}(0)(1-\Prob_{S}(l))=\Prob_{M'}(0)(1-\frac{1}{N})$, where $S$ denotes the random server to which the $\#_{\theta}$ query is sent. Thus, we have
        \begin{align}
        I(\theta; Q_{l}^{\theta}) 
        \overset{}{=}& \log M - \biggl(\Prob_{M'}(0)\left(1-\frac{1}{N}\right)\log M\nonumber\\ \label{eqn:mileakagekeystep}&+ \sum_{m' = 1}^{M-1}\log(m'+1) \biggl(\sum_{\substack{\pmb{q}\in \mathcal{Q}_l:\\|{\cal M}_{\pmb{q}}|=m'+1}}\Prob_{Q_{l}^{\theta}}(\pmb{q})\biggl) \biggl)\\
        \overset{(b)}{=}& \log M - \biggl(\Prob_{M'}(0)\bigg(1-\frac{1}{N}\bigg) \log M\nonumber\\ &~~~~~~~~~~~~~~~~~~~~+ \sum_{m' = 1}^{M-1}\log(m'+1) \Prob_{M'}(m') \biggl)\nonumber\\
        \label{eqn:mileakagewsjqed}
        \overset{}{=}& \biggl(1 - \bigg(1-\frac{1}{N}\bigg)\Prob_{M'}(0)\biggl) \log M- \expec[\log(M'+1)],
    \end{align}
where $(b)$ holds as the sum of all set of queries which contain $m'+1$ messages is exactly $\Prob_{M'}(m')$. This completes the proof.
\end{IEEEproof}
\begin{lemma}
\label{lemma:WSJ_MaxL_Leakage}
    \begin{align}
    \nonumber
       &\maxleak(\wsj)= \log M + \log\biggl(\expec\biggl[\frac{1}{M'+1}\biggl]\\
       \label{eqn:maxleakagemetric_WSJ}
       &\hspace{2.5cm}-\bigg(1-\frac{1}{M}\bigg)\bigg(1-\frac{1}{N}\bigg)\Prob_{M'}(0)\biggl).
    \end{align}
\end{lemma}
\begin{IEEEproof}
We first make a few observations. Firstly, $\Prob_{Q_{l}^{\theta}|\theta}(\pmb{q}|m)=$ $\Prob_{\theta|Q_{l}^{\theta}}(m|\pmb{q})\Prob_{Q_{l}^{\theta}}(\pmb{q})/\Prob_{\theta}(m)=M\Prob_{\theta|Q_{l}^{\theta}}(m|\pmb{q})\Prob_{Q_{l}^{\theta}}(\pmb{q}).$ Further, by Claim \ref{claim:thetagivenQ}, $\max_{m\in [1:M]}\Prob_{\theta|Q_{l}^{\theta}}(m|\#_k)=\Prob_{\theta|Q_{l}^{\theta}}(k|\#_k)=1$ and $\max_{m\in [1:M]}\Prob_{\theta|Q_{l}^{\theta}}(m|\phi)=1/M$. Also by Claim \ref{claim:thetagivenQ}, when $|{\cal M}_{\pmb{q}}|=m'+1\geq 2$, then $\max_{m\in [1:M]}\Prob_{\theta|Q_{l}^{\theta}}(m|\pmb{q})=1/(m'+1)$. Using these, we now show that the leakage $\mathsf{MaxL}(\theta,Q_l^{\theta})$  at each server $l\in[0:N-1]$ has the expression equal to the R.H.S. of \eqref{eqn:maxleakagemetric_WSJ}, which completes the proof. 
    \begin{align}
        \mathsf{MaxL}&(\theta,Q_l^{\theta})\nonumber \\
        \overset{}{=}&\log \biggl(\sum_{\pmb{q}\in \mathcal{Q}_l} \max_{m\in [1:M]} \Prob_{Q_{l}^{\theta}|\theta}(\pmb{q}|m)\bigg) \nonumber\\
        \overset{}{=}&\log \biggl(\sum_{k=1}^{M} M\Prob_{Q_{l}^{\theta}}(\#_k)\max_{m\in [1:M]}\Prob_{\theta|Q_{l}^{\theta}}(m|\#_k)\nonumber\\ &+ M\Prob_{Q_{l}^{\theta}}(\phi)\max_{m\in [1:M]}\Prob_{\theta|Q_{l}^{\theta}}(m|\phi)\nonumber\\ &+ \sum_{m'=1}^{M-1}  \sum_{\substack{\pmb{q}\in \mathcal{Q}_l:\\ |{\cal M}_{\pmb{q}}| = m'+1}} \hspace{-0.3cm}M\Prob_{Q_{l}^{\theta}}(\pmb{q})\max_{m\in [1:M]}\Prob_{\theta|Q_{l}^{\theta}}(m|\pmb{q})\biggl)\label{eqn:maxleakage_intermed_step}\\
        \label{eqn:maxleakagekeystep}
        \overset{}{=}&\log\bigg(M\biggl(\sum_{k=1}^{M} \Prob_{\theta}(k)\Prob_{S}(l)\Prob_{M'}(0)\cdot 1\nonumber\\ &~~~~~~~~~~ +(1-\Prob_{S}(l))\Prob_{M'}(0)\cdot \frac{1}{M}\nonumber\\ &~~~~~~~~~+ \sum_{m'=1}^{M-1} \frac{1}{m'+1} \sum_{\substack{\pmb{q}\in \mathcal{Q}_l: \\|{\cal M}_{\pmb{q}}| = m'+1}} \Prob_{Q_{l}^{\theta}}(\pmb{q}) \biggl)\bigg)\\
        \overset{}{=}&\log \bigg(M\biggl(\frac{\Prob_{M'}(0)}{N} + \frac{(N-1)\Prob_{M'}(0)}{MN}\nonumber\\ &~~~~~~~~~~~~~~~~+ \sum_{m'=1}^{M-1}\frac{1}{m'+1}\Prob_{M'}(m') \biggl)\bigg)\nonumber\\
        \overset{}{=}& \log M + \log\biggl(\expec\biggl[\frac{1}{M'+1}\biggl]\nonumber\\&- \bigg(1-\frac{1}{M}\bigg)\bigg(1-\frac{1}{N}\bigg)\Prob_{M'}(0)\biggl).\nonumber
    \end{align}
\end{IEEEproof}
\subsection{Achievable rate-privacy trade-offs with specific $\Prob_{M'}$}
Using Lemmas \ref{lemma:WSJ_MI_Leakage} and \ref{lemma:WSJ_MaxL_Leakage}, we now present achievable rate-privacy trade-offs for collusion-free replicated-servers WPIR in Theorem \ref{thm:WSJ_MIL} and \ref{thm:WSJ_MaxL}. These are obtained by time-sharing the Sun-Jafar protocol (with all $M$ files) and a clean download phase, while avoiding any other phase with $M'\notin\{0,M-1\}$.
\begin{theorem}
    \label{thm:WSJ_MIL}
    For any $\rho\geq 0$, the protocol $\wsj$ achieves the following rate-privacy trade-off.
    \begin{itemize}
        \item $\mileak=\min\left(\rho,\frac{\log M}{N}\right)$,
        \item $R=\biggl(1+\left(1-\frac{\rho N}{\log{M}}\right)_+\left(\frac{1}{N} + \cdots + \frac{1}{N^{M-1}}\right)\biggl)^{-1}$,
    \end{itemize}
    where $(x)_{+} \triangleq \max(x,0)$.
\end{theorem}

\begin{IEEEproof}
If $\rho\geq\frac{\log M}{N}$, then the client can choose the probability distribution $\Prob_{M'}$ in the protocol $\wsj$ as $\Prob_{M'}(0)=1$ and $\Prob_{M'}(m')=0, \forall m'\neq 0$. It then follows from \eqref{eqn:rategeneralpm'} that the rate is $1$, and from Lemma \ref{lemma:WSJ_MI_Leakage} that $\mileak=\frac{\log M}{N}$. 

Now, consider $\rho\in[0,\frac{\log M}{N}]$. The client can then choose $\Prob_{M'}(0)=\frac{\rho N}{\log M}$, and $\Prob_{M'}(M-1)=1-\frac{\rho N}{\log M}$. Now, using \eqref{eqn:rategeneralpm'} the rate of this scheme is: $R = (1-\frac{1}{N})/(1 - \frac{\Prob_{M'}(0)}{N} - \frac{\Prob_{M'}(M-1)}{N^M}) = (1 + (1-\Prob_{M'}(0))(\frac{1}{N} + \cdots + \frac{1}{N^{M-1}}))^{-1}$. Using the value of $\Prob_{M'}(0)$ considered, we get the expression for the rate as given in the statement. Finally, we calculate $\mileak$. We see that $\expec[\log(M'+1)]=\Prob_{M'}(M-1)\log(M)$. Plugging this in Lemma \ref{lemma:WSJ_MI_Leakage}, we see that $\mileak = \rho$, for the chosen distribution $\Prob_{M'}$
. This completes the proof. 

\end{IEEEproof}
\begin{theorem}
    \label{thm:WSJ_MaxL}
    For any $\rho \geq 0$, the protocol $\wsj$ achieves the following rate and MaxL leakage pairs
    \begin{itemize}
        \item $\maxleak = \min\left(\rho, \log\left(1+\frac{M-1}{N}\right)\right)$,
        \item $R = \biggl(1+\left(1-N\frac{2^{\rho}-1}{M-1}\right)_{+}\left(\frac{1}{N} + \cdots + \frac{1}{N^{M-1}}\right)\biggl)^{-1}$.
    \end{itemize}
\end{theorem}
\begin{IEEEproof}
If $\rho\geq\log (1+\frac{M-1}{N})$, then the client chooses the distribution $\Prob_{M'}$ in the protocol $\wsj$ as $\Prob_{M'}(0) = 1$ and $\Prob_{M'}(m') = 0. \forall m' \in [1:M-1]$. Using \eqref{eqn:rategeneralpm'} the rate is $1$, and from Lemma \ref{lemma:WSJ_MaxL_Leakage} that $\maxleak = \log(1+\frac{M-1}{N})$. 

Now, consider $\rho\in \left[0,\log(1+\frac{M-1}{N})\right]$. The client can choose $\Prob_{M'}(0)=N\frac{2^{\rho}-1}{M-1}$, and $\Prob_{M'}(M-1) = 1 - \Prob_{M'}(0)$.
Now, using \eqref{eqn:rategeneralpm'}, the rate will be: $R = (1-\frac{1}{N})/(1 - \frac{\Prob_{M'}(0)}{N} - \frac{\Prob_{M'}(M-1)}{N^M}) = (1 + (1-\Prob_{M'}(0))(\frac{1}{N} + \cdots + \frac{1}{N^{M-1}}))^{-1}$. Plugging the value of $\Prob_{M'}(0) = N\left(\frac{2^{\rho}-1}{M-1}\right)$, we get the rate as stated in this theorem.
Using the fact that $\expec[\frac{1}{M'+1}]=\Prob_{M'}(0)+\frac{\Prob_{M'}(M-1)}{M+1}$ and Lemma \ref{lemma:WSJ_MaxL_Leakage}, we see that the $\maxleak = \rho$ for the distribution considered. This completes the proof.  
\end{IEEEproof}

\begin{figure}[htbp]
    \centering
    \includegraphics[width=0.5\textwidth]{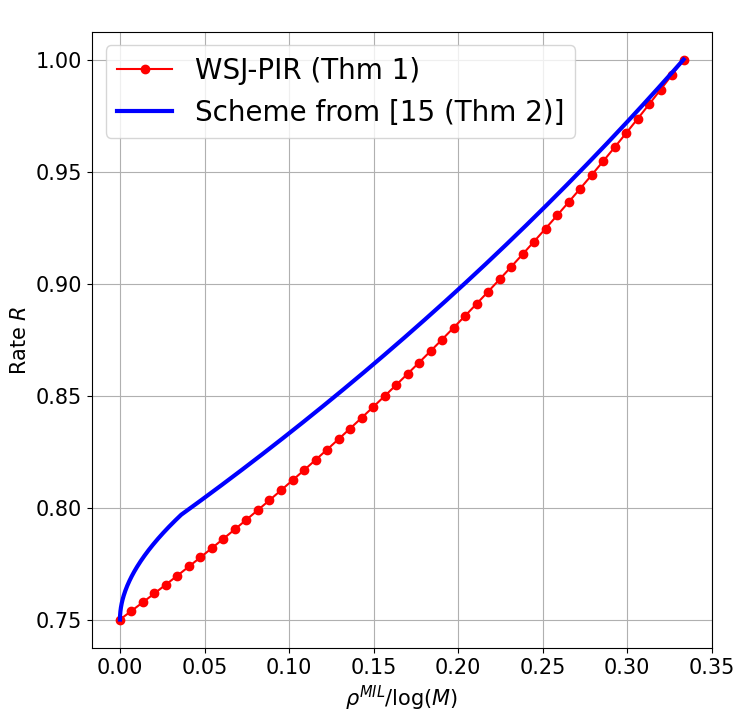}
        \caption{Rate vs normalized $\mileak$ comparison between $\wsj$ (contiguous, blue) and scheme from \cite{improved2022wpir} (dotted, red) with $N=3$, $M=2$.} 
    \label{fig:comparisonmileak_nocollusion}
\end{figure}

\begin{remark}
    While we do not have a proof for the optimality of the chosen distributions in Theorems \ref{thm:WSJ_MIL} and \ref{thm:WSJ_MaxL}, we have numerically observed that these distributions result in the best rate-privacy trade-offs among many natural choices for $\Prob_{M'}$. Further, the trade-off under maximal leakage in Theorem \ref{thm:WSJ_MaxL} is identical with the trade-off reported in \cite[Theorem 1]{improved2022wpir}, which is the best known trade-off so far under the maximal leakage metric, as per our knowledge. On the other hand, the trade-off given in \cite[Theorem 2]{improved2022wpir} for the mutual information metric is better than that in Theorem \ref{thm:WSJ_MIL}. We illustrate this gap for the case of $M=2, N=3$ in Fig. \ref{fig:comparisonmileak_nocollusion}. However, we have observed that this gap becomes negligible, as the values of $N$ and $M$ increase.
\end{remark}

\section{Extensions to MDS \& T-collusion WPIR}
\label{sec:WPIR-collusion-MDS-Tcoll}

In this section, we present WPIR protocols for the MDS-coded servers without collusion and for $T$-colluding servers under replicated storage. Essentially, these are extensions of the respective capacity-achieving protocols in \cite{banawan2018capacity} and \cite{sun2017capacityb}, in the same way as the protocol $\wsj$ is an extension of the Sun-Jafar protocol. Particularly, the simple overlay of a clean download phase and the time-sharing trick results in our $T$-collusion WPIR protocol, which, to the best of our knowledge, is the first one for this setting. In the case of MDS-coded storage, we compare our trade-offs with those in a very recent work \cite{orvedal2024weaklyprivateinformationretrievalmdscoded}, which is based on ideas in the TSC scheme \cite{tian2019capacity,Zhouetal_2019_TIT_MDS_optimalmsgsize_lowerupcost,Zhuetal_2020_MDS_optimalmessagesize_higher_upcost}.
\subsection{A new WPIR scheme for MDS-coded servers}
\label{subsec:MDScodedWPIR}
Banawan and Ulukus \cite{banawan2018capacity} presented a PIR protocol that achieves the capacity (given in \eqref{eqn:MDScapacity}) for the system consisting of $N$ servers storing the MDS-coded symbols of the $M$ files. The scheme in \cite{banawan2018capacity}, which we denote as $\banuluk$, requires the files to be viewed as matrices of size $N^M\times K$, which are encoded using a $(N,K)$-MDS code into an $N^M\times N$ matrix of encoded symbols, and placed in $N$ servers column-wise. Subsequently, the protocol $\banuluk$ works along the lines of $\sunjaf$ for $K$ iterations, to recover sufficient encoded subfiles of the desired file to decode the same. 


Lifting ideas from Section \ref{sec:sjwpircollusionfree}, we immediately see the possibility of a WPIR scheme for the MDS-coded system. We denote this new protocol $\wbu$. We also need the super-segmentation idea from Section \ref{sec:sjwpircollusionfree}. For each $s\in[2:M]$, the $s$-level super-segmentation of each encoded file involves partitioning its $N^M$ rows of the file into $N^s$ super-rows (each of length $N$), where each super-row consists of a distinct subset of $N^{M-s}$ original rows. 

\textit{Query Generation of $\wbu$:} Let the desired file index be $\theta = k$. As in Section \ref{sec:sjwpircollusionfree}, we assume the client chooses the value of the random variable $M'$ as per a chosen distribution $\Prob_{M'}$ on  $[0:M-1]$. The query at server $l$ is generated as follows. 
    \begin{itemize} 
        \item \textit{Case 1:} If $M'=0$, the client selects a subset $S$ of $K$ distinct servers indices, uniformly at random from all possible $K$-subsets of the $[0:N-1]$. The client sends query $\#_k$ to the $K$ servers indexed by $S$ and sends query $\phi$ (null query) to all other servers.
        \item \textit{Case 2:} If $M'=m\neq 0$, the client selects $m'$ distinct indices of non-desired files uniformly at random, denoted by $J\subset [1:M]\setminus \{k\}$. The client then executes the query-generation protocol as per the protocol ${\cal P}_{\mathsf{BU}}^{(m'+1)}$ on the set of files $\{W_j, j\in J\} \cup \{W_k\}$ and the desired file index $k$, and using $(m'+1)$-level super-segmentation.
    \end{itemize}

\textit{Server Responses:} 
    \begin{itemize}
        \item If $Q_l = \#_k$, server $l$ sends the encoded part of $W_k$ which is available in that server. 
        \item If $Q_l = \phi$, server $l$ sends no response.
        \item For any $Q_l=\pmb{q}\notin \{\phi\}\cup\{\#_m:m\in[1:M]\}$, 
        the server responds according to the $\bumprime$ protocol, on files with indices 
        $J \cup \{k\}$ (this set can be identified from $\pmb{q}$), and $|{\cal M}_{\pmb{q}}|$-level super-segmentation, where ${\cal M}_{\pmb{q}}$ refers to the set of file indices which appear in the query $\pmb{q}$.
    \end{itemize}

\subsubsection{Correctness and Rate}
Suppose $M'=0$. Then, by the MDS property, the client can decode the desired file $W_k$ from the $K$ distinct encoded parts obtained from the servers $S$. 
When $M'\neq 0$, the correctness of the protocol $\wbu$ follows directly from the arguments proved in \cite{banawan2018capacity}.

Recalling that the protocol $\banuluk$ has the rate as in \eqref{eqn:MDScapacity}, the rate of the protocol $\wbu$ can be derived similar to $\wsj$, with only appropriate changes. We omit the simple calculation and give only the final expression obtained here, as follows. 
\begin{align}
\label{eqn:ratewbuprotocol}
    R_{\mathsf{WBU}} = \frac{1-\frac{K}{N}}{1-\expec\left[{\left(\frac{K}{N} \right)}^{M'+1}\right]},
\end{align}
where the expectation is with respect to the distribution of $M'$.
\subsubsection{Privacy metrics for arbitrary $\Prob_{M'}$}
To derive the privacy metrics for $\wbu$ protocol for some chosen $\Prob_{M'}$, we follow the same steps as outlined in Lemmas \ref{lemma:WSJ_MI_Leakage} and \ref{lemma:WSJ_MaxL_Leakage} for protocol $\wsj$. We summarize only the changes and then present the results without elaborate proofs. Note that, in $\wbu$, the set $S$ is a random subset of $K$ distinct server indices. Thus, for each $l\in[0:N-1]$, we have $\Pr(l\in S)=\binom{N-1}{K-1}/\binom{N}{K} = \frac{K}{N}$. Hence, we have $\Prob_{Q_l^{\theta}}(\phi)=\Prob_{M'}(0)(1-\Pr(l\in S))=\Prob_{M'}(0)(1-K/N)$ and $\Prob_{Q_l^{\theta}}(\#_m)=\Prob_{M'}(0)\Pr(l\in S)=\Prob_{M'}(0)\frac{K}{N}$. Further, Claim \ref{claim:thetagivenQ} can be shown to be true in this setup also.  Effecting these changes at steps \eqref{eqn:mileakcalculation_intermed_sj} and \eqref{eqn:maxleakage_intermed_step} in the calculations of $\mileak$ and $\maxleak$ respectively, and following the rest of the arguments, we obtain the following results. 

\begin{lemma} 
\label{lemma:MDS_WSJ_MI_leakage}
    \begin{align}
        \mileak(\wbu) &= \log M\left[1 - \left( 1 - \frac{K}{N} \right)P_{M'}(0) \right]\nonumber\\ &- \expec\left[ \log(M'+1) \right]
    \end{align}
\end{lemma}

\begin{lemma} 
\label{lemma:MDS_WSJ_MaxL_leakage}
    \begin{align}
        \maxleak(\wbu) &= \log M + \log \Biggl( \expec \left[\frac{1}{M'+1}\right] \nonumber\\ &- \left(1-\frac{1}{M}\right)\left(1- \frac{K}{N}\right)P_{M'}(0)\Biggl)
    \end{align}
\end{lemma}
The theorems below follow similarly to Theorems \ref{thm:WSJ_MIL} and \ref{thm:WSJ_MaxL}. We choose the distribution $\Prob_{M'}(0)=\min(\frac{\rho N}{K\log M},1)=1-\Prob_{M'}(M-1)$ in Lemma \ref{lemma:MDS_WSJ_MI_leakage} for proving Theorem \ref{thm:WBU_MIL}, while for Theorem \ref{thm:WBU_MaxL}, we choose $\Prob_{M'}(0)=\min(\frac{(2^{\rho}-1) N}{K(M-1)},1)=1-\Prob_{M'}(M-1)$ in Lemma \ref{lemma:MDS_WSJ_MaxL_leakage}. The rest of the proofs are as before, and hence omitted. 
\begin{theorem}
    \label{thm:WBU_MIL}
    For any $\rho\geq 0$, the protocol $\wbu$ achieves the following rate-privacy trade-off under mutual information leakage.
    \begin{itemize}
        \item $\mileak = \min(\rho, \frac{K\log M}{N})$,
        \item $R = \left(1+ \left(1-\frac{N\rho}{K\log M}\right)_{+} \left(\frac{K}{N} + \cdots + \left(\frac{K}{N}\right)^{M-1} \right)\right)^{-1}$.
    \end{itemize}
\end{theorem}
\begin{theorem}
    \label{thm:WBU_MaxL}
    For any $\rho\geq 0$, the protocol $\wbu$ achieves the following rate-privacy trade-off under maximal leakage. 
    \begin{itemize}
        \item $\maxleak = \min(\rho, \log (1 + \frac{K(M-1)}{N}))$,
        \item $R = \left(1+ \left(1-\frac{N(2^\rho - 1)}{K(M-1)}\right)_{+} \left(\frac{K}{N} + \cdots + \left(\frac{K}{N}\right)^{M-1} \right)\right)^{-1}$.
    \end{itemize}
\end{theorem}
     
\begin{figure}[htbp]
    \includegraphics[width=0.5\textwidth]{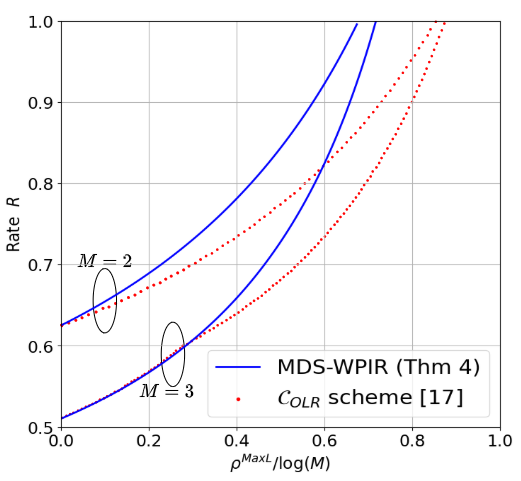}
    \caption{Rate vs. normalized $\maxleak$ leakage trade-off comparison between $\wbu$ (contiguous, blue) and scheme $\mathcal{C}_{OLR}$ \cite{orvedal2024weaklyprivateinformationretrievalmdscoded} (dotted, red) for $(5,3)$-MDS-coded storage with $M\in\{2,3\}$.}
    \label{fig:MDS_MaxL1} 
\end{figure}
\begin{figure}[htbp]
    \includegraphics[width=0.5\textwidth]{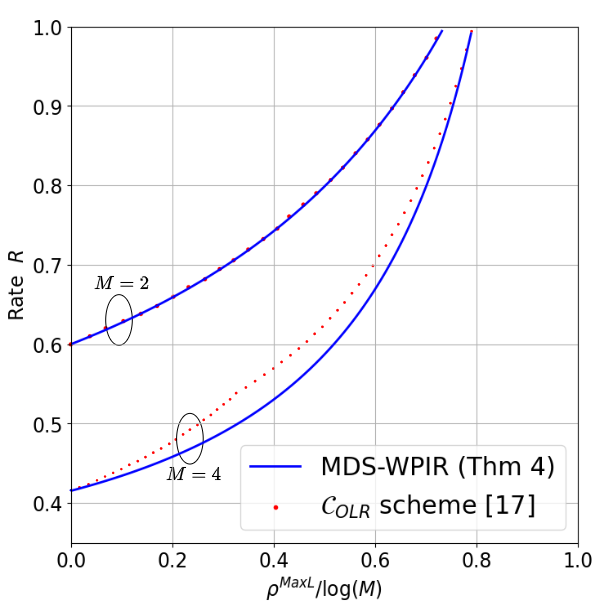}
    \caption{Rate vs. normalized $\maxleak$ leakage trade-off comparison between $\wbu$ (contiguous, blue) and scheme $\mathcal{C}_{OLR}$ \cite{orvedal2024weaklyprivateinformationretrievalmdscoded} (dotted, red) for $(3,2)$-MDS-coded storage with $M=2$, and $M=4$.}
    \label{fig:MDS_MaxL2}
\end{figure}
\begin{figure}[htbp]
    \includegraphics[width=0.5\textwidth]{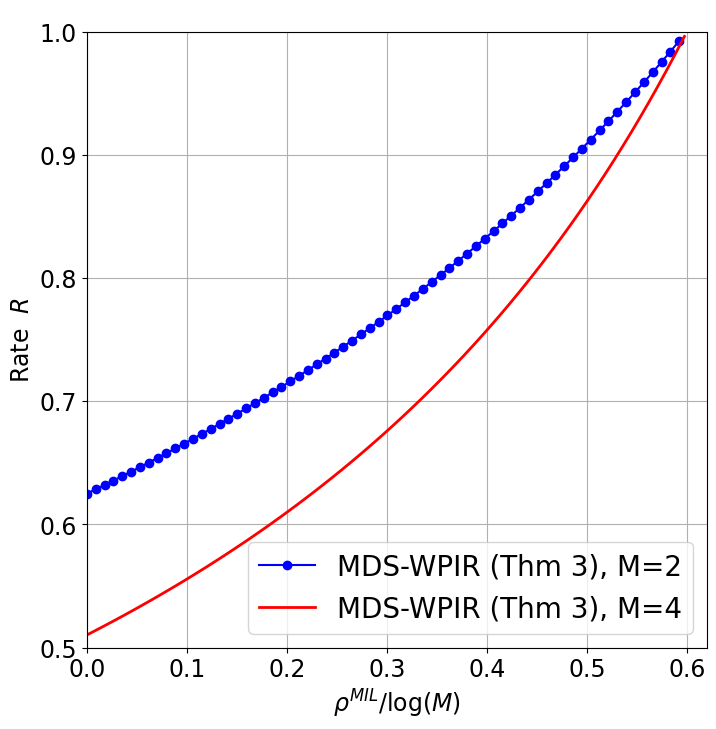}
    \caption{Trade-off between rate and normalized $\mileak$ leakage for $\wbu$ for $(5,3)$-MDS-coded storage with $M=2$ (blue dotted curve), $M=3$ (red, contiguous curve).}
    \label{fig:MDS_MIL} 
\end{figure}

To the best of our knowledge, the only other known WPIR schemes for the MDS-coded setup are the three schemes shown in \cite{orvedal2024weaklyprivateinformationretrievalmdscoded}. Explicit expressions for the trade-offs achieved by these schemes are not available in \cite{orvedal2024weaklyprivateinformationretrievalmdscoded} for all parameters; however, the authors provide numerically solved trade-offs under maximal leakage for some examples. In particular, the `OLR' scheme from \cite[Section IV]{orvedal2024weaklyprivateinformationretrievalmdscoded} achieves the best trade-off known so far (for the entire regime sweeping for rates ranging from $1$ to zero-leakage capacity \eqref{eqn:MDScapacity}), for the settings of $(N=3, K=2)$ and $(N=5,K=3)$ under maximal leakage. Fig. \ref{fig:MDS_MaxL1} and Fig. \ref{fig:MDS_MaxL2} shows a comparison between the presented trade-off in Theorem 
\ref{thm:WBU_MaxL}, and the `OLR' scheme for these parameters. We observe that our scheme from Theorem \ref{thm:WBU_MaxL} has superior performance than the OLR scheme under maximal leakage, for parameters $(N=5, K=3)$; while for parameters $(N=3, K=2)$, our scheme matches the OLR scheme for $M=2$, but has slightly inferior performance when $M=4$. We suspect that our scheme will perform better for larger values of $N$ and $K$ in general, though we cannot confirm this at this juncture, as the optimization of the OLR scheme becomes tedious for larger parameters. 
To the best of our knowledge, there is no known trade-off between rate and the $\mileak$ leakage metric in the existing literature. Thus, for the case of mutual information leakage, our trade-off
appears to be the first known in literature. Using Theorem \ref{thm:WBU_MIL}, we can plot the rate-privacy ($\mileak$) trade-off. Fig. \ref{fig:MDS_MIL} shows the rate-privacy ($\mileak$) trade-off curves for the setting of ($N=5$, $K=3$) with $M=2$ and $M=4$. 

\subsection{First WPIR scheme for $T$-collusion}
\label{subsec:Tcoll-WPIR}

In the $T$-collusion setting, a group of up to $T$ servers collude, and it is essential to ensure privacy is preserved even among the colluding servers. The WPIR protocol we present is built upon the capacity-achieving PIR protocol from \cite{sun2017capacityb}, using ideas as in Section \ref{sec:sjwpircollusionfree}. We denote this protocol as $\wcsj$, which to the best of our knowledge is the first in this setting. We now describe the query generation of $\wcsj$.


\textit{Queries, Server responses and Correctness of $\wcsj$:} We have $L=N^M$ as the number of segments in each file, and these segments have symbols from a sufficiently large field, as required in \cite{sun2017capacityb}. We reuse the super-segmentation idea as in Sub-section \ref{subsec:WPIRSJ_queryserverresponses}. The query generation is almost identical as in the protocol $\wsj$. In particular, the generation of $M'$ and Case 1 is identical to that of $\wsj$. In Case 2, we invoke the $T$-collusion protocol from \cite{sun2017capacityb} for the $M'$ distinct non-desired files (chosen uniformly at random) and the desired file, where $M'$ is chosen according to some distribution $\Prob_{M'}$. The server responses too are similar to those in $\wsj$ (a) we have a clean download from a random server if the randomly chosen value of $M'$ is $0$, and (b) in case of $Q_l^{\theta}=\pmb{q}_l\notin \{\phi\}\cup\{\#_m:m\in[1:M]\}$ (which happens when $M'\neq 0$), the server $l$ responds as per the protocol in \cite{sun2017capacityb} using super-segments of the desired file along with $M'$ randomly chosen non-desired subfiles. 
The correctness of the $\wcsj$ protocol directly follows the arguments proved in \cite{sun2017capacityb} when $M'\neq 0$, while for $M'=0$ we have the clean download of the desired file $W_k$ from a server, which is obviously correct. 

\subsection{Rate and Privacy of $\collsunjaf$}


From \eqref{eqn:collcapacity} and the description of protocol $\wcsj$, we have that the total download from all servers is of size $N^{M'+1}(1+\frac{T}{N}+\cdots+(\frac{T}{N})^{M'})N^{M-(M'+1)}$, for all values of $M'\in[0:M-1]$. Using this observation and the fact that the file size is $L=N^M$, the rate expression below follows directly from the definition.
\begin{align}
\label{eqn:rate_Tcollusion_WPIR}
    R_{T-\mathsf{WSJ}} = \frac{1-\frac{T}{N}}{1-\expec\left[{\left(\frac{T}{N} \right)}^{M'+1}\right]},
\end{align}
where the expectation is over the distribution of $M'$.

To compute the privacy metrics for some chosen distribution $\Prob_{M'}$, we need the equivalent of Claim \ref{claim:thetagivenQ}. We need a few notations to present this claim, which we give now. 
For ${\cal T}\in\binom{[0:N-1]}{T}$, let $Q_{\cal T}^{\theta}=(Q_t^{\theta}: t\in {\cal T})$ denote the $T$-tuple of random queries seen by the servers indexed by ${\cal T}$, and $\pmb{q}_{\cal T}=(\pmb{q}_t: t\in {\cal T})$ denote an arbitrary realization of the same. For some query $\pmb{q}$ in the query set of any server seen in the protocol $\wcsj$, we say $\pmb{q}\in\pmb{q}_{\cal T}$ if $\pmb{q}_t=\pmb{q},$ for some $t\in{\cal T}$. For any $\pmb{q}_{\cal T}$ being a $T$-tuple queries seen by servers $\cal T$ occurring in an instantiation of the $T$-collusion PIR protocol in \cite{sun2017capacityb} on $M'+1$ files, let ${\cal M}_{\pmb{q}_{\cal T}}$ denote the set of file indices that appear in the tuple of queries $\pmb{q}_{\cal T}$. In this case, we have that ${\cal M}_{\pmb{q}_{\cal T}}=M'+1$, which is known from the query structure in \cite{sun2017capacityb}. Now, we present the claim. We omit the proof, as it follows almost exactly on the lines of Claim \ref{claim:thetagivenQ}.
\begin{claim}
    \label{claim:thetagivenQ_Tcoll}
    With respect to the protocol $\wcsj$, the following statements are true for any $m\in[1:M]$. 
    \begin{enumerate}
        \item For any $k\in[1:M]$, $$\Prob_{\theta|Q_{\cal T}^{\theta}}(m|\pmb{q}_{\cal T})=\begin{cases}1&\text{if}~\#_m\in \pmb{q}_{\cal T}\\
        0&\text{if}~\#_k\in \pmb{q}_{\cal T}, \text{where}~m\neq k.\end{cases}$$
        \item If $\pmb{q}_{\cal T}=\overbrace{(\phi,\phi,\hdots,\phi)}^{T \text{times}}$, then $\Prob_{\theta|Q_{\cal T}^{\theta}}(m|\pmb{q}_{\cal T})=\frac{1}{M}$. 
        \item If $\phi\notin \pmb{q}_{\cal T}$ and $\#_k\notin \pmb{q}_{\cal T}$ for any $k\in[1:M]$, then 
    $$\Prob_{\theta|Q_{\cal T}^{\theta}}(m|\pmb{q}_{\cal T})=\begin{cases}\frac{1}{|{\cal M}_{\pmb{q}_{\cal T}}|}, &\text{if}~m\in {\cal M}_{\pmb{q}_{\cal T}},\\
        ~~~0,&\text{if}~m\notin {\cal M}_{\pmb{q}_{\cal T}},\end{cases}$$
    \end{enumerate}
    
\end{claim}

As in Section \ref{sec:sjwpircollusionfree}, we now derive the metrics $\mileak$ and $\maxleak$ for $\wcsj$ protocol. To facilitate this, we make a few observations. If $M'=0$, then recall that $S$ (as per Case 1 of the query generation) denotes the server index to which the $\#_{\theta}$ query is sent. For a subset $\cal T$ of $T$ servers, we thus have $\Pr(S\in {\cal T})=\frac{T}{N}$. Using Claim \ref{claim:thetagivenQ_Tcoll}, we obtain our privacy metrics in the following lemmas. Their proofs are on the lines of the proofs of Lemmas \ref{lemma:WSJ_MI_Leakage} and \ref{lemma:WSJ_MaxL_Leakage}, with appropriate changes.


\begin{lemma} 
\label{lemma:TWSJ_MI_leakage}
\begin{align}
    \mileak(\wcsj) &= \log M\left[1 - \left( 1 - \frac{T}{N} \right)P_{M'}(0) \right]\nonumber\\ &~~~~~~~~~~~~~~~~~~ - \label{eqn:lemma5}\expec\left[ \log(M'+1) \right].
\end{align}
\end{lemma}
\begin{IEEEproof}
It is sufficient to show $I(\theta; Q_{\mathcal{T}}^{\theta})$ is equal to the R.H.S of \eqref{eqn:lemma5} for any subset $\mathcal{T}\subset [0:N-1]$ of size $T$. Using \eqref{eqn:MI_leakage_definition_Tcoll}, the proof will follow.
\begin{align}
    I(\theta; &Q_{\mathcal{T}}^{\theta})\nonumber\\ 
    \overset{}{=}& H(\theta) - H(\theta | Q_{\mathcal{T}}^{\theta})\nonumber\\
    \overset{}{=}& \log M - \sum_{\pmb{q}_{\mathcal{T}} \in \mathcal{Q}_{\mathcal{T}}} \Prob_{Q_{\mathcal{T}}^{\theta}}(\pmb{q}_{\mathcal{T}})H(\theta | Q_{\mathcal{T}}^{\theta}=\pmb{q}_{\mathcal{T}})\nonumber\\
    \overset{}{=}& \log M - \biggl(\sum_{k=1}^{M}\sum_{\substack{\pmb{q}_{\mathcal{T}} \in \mathcal{Q}_{\mathcal{T}}:\nonumber\\ \#_k \in \pmb{q}_{\mathcal{T}}}} \Prob_{Q_{\mathcal{T}}^{\theta}}(\pmb{q}_{\mathcal{T}})H(\theta | Q_{\mathcal{T}}^{\theta}=\pmb{q}_{\mathcal{T}})\nonumber\\
    &~~~~~~~~~+ \sum_{\substack{\pmb{q}_{\mathcal{T}} \in \mathcal{Q}_{\mathcal{T}}:\nonumber\\ \pmb{q}_{\mathcal{T}}=(\phi,\phi,\hdots,\phi)}} \Prob_{Q_{\mathcal{T}}^{\theta}}(\pmb{q}_{\mathcal{T}})H(\theta | Q_{\mathcal{T}}^{\theta}=\pmb{q}_{\mathcal{T}})\nonumber\\
    &~~~~~~~~~+ \sum_{m' = 1}^{M-1} \sum_{\substack{\pmb{q}_\mathcal{T}\in \mathcal{Q}_{\mathcal{T}}: \\|{\cal M}_{\pmb{q}_\mathcal{T}}|=m'+1}}\Prob_{Q_{\mathcal{T}}^{\theta}}(\pmb{q}_\mathcal{T})H(\theta|Q_{\mathcal{T}}^{\theta}=\pmb{q}_\mathcal{T}) \biggl)\label{eqn:proofMIL_TSJ:stage1}
    \end{align}
    Using Claim \ref{claim:thetagivenQ_Tcoll}, it is easy to the following. If $\#_k \in \pmb{q}_{\mathcal{T}}$, then $H(\theta | Q_{\mathcal{T}}^{\theta}=\pmb{q}_{\mathcal{T}}) = 0$. Also, if $\pmb{q}_{\mathcal{T}}=(\phi, \phi, \cdots,\phi)$, then clearly $H(\theta | Q_{\mathcal{T}}^{\theta}=\pmb{q}_{\mathcal{T}}) = \log M$ and $\Prob_{Q_{\mathcal{T}}^{\theta}}(\pmb{q}_\mathcal{T}) = \Prob_{M'}(0)\Pr(S\notin \mathcal{T})= \Prob_{M'}(0)(1-\frac{T}{N})$. Finally, 
    If $|\mathcal{M}_{\pmb{q}_{\mathcal{T}}}|>1$, $H(\theta | Q_{\mathcal{T}}^{\theta}=\pmb{q}_{\mathcal{T}}) = \log|\mathcal{M}_{\pmb{q}_{\mathcal{T}}}|$. Using these in \eqref{eqn:proofMIL_TSJ:stage1}, we have
    \begin{align}
    I(\theta; Q_{\mathcal{T}}^{\theta}) 
    \overset{}{=}& \log M - \biggl(\Prob_{M'}(0)\left(1-\frac{T}{N}\right)\log M\nonumber\\ &+ \sum_{m' = 1}^{M-1}\log(m'+1) \biggl(\sum_{\substack{\pmb{q}_{\mathcal{T}}\in \mathcal{Q}_{\mathcal{T}}:\nonumber\\|{\cal M}_{\pmb{q}_{\mathcal{T}}}|=m'+1}}\Prob_{Q_{\mathcal{T}}^{\theta}}(\pmb{q}_{\mathcal{T}})\biggl) \biggl)\\
    \overset{(a)}{=}& \log M - \biggl(\Prob_{M'}(0)\bigg(1-\frac{T}{N}\bigg) \log M\nonumber\\ &~~~~~~~~~~~~~~~~~~~~+ \sum_{m' = 1}^{M-1}\log(m'+1) \Prob_{M'}(m') \biggl)\nonumber\\
    \label{eqn:miTleakagewsjqed}
    \overset{}{=}& \biggl(1 - \bigg(1-\frac{T}{N}\bigg)\Prob_{M'}(0)\biggl) \log M- \expec[\log(M'+1)],
    \end{align}
    where $(a)$ holds as the sum of all $T$-tuples of queries to ${\cal T}$ which contain $m'+1$ messages is exactly $\Prob_{M'}(m')$, by Case 2 of the protocol. This completes the proof. 
\end{IEEEproof}

\begin{lemma} 
\label{lemma:TWSJ_Max_leakage}
    \begin{align}
         \maxleak(\wcsj) &= \log M + \log\biggl(\expec\left[\frac{1}{M'+1}\right] \nonumber\\ \label{eqn:lemma6statement} &- \left(1-\frac{1}{M}\right)\left(1- \frac{T}{N}\right)P_{M'}(0) \biggl).
    \end{align}
\end{lemma}
\begin{IEEEproof}
    As in the proof of Lemma \ref{lemma:WSJ_MaxL_Leakage}, it is enough to show that $\mathsf{MaxL}(\theta, Q_{\mathcal{T}}^{\theta})$ is equal to the R.H.S of \eqref{eqn:lemma6statement}, for any ${\cal T}\in\binom{[0:N-1]}{T}$. We have the following.
    \begin{align}
        \mathsf{MaxL}(\theta, &Q_{\mathcal{T}}) \nonumber\\
        &\overset{}{=} \log \biggl(\sum_{\pmb{q}_{\mathcal{T}}\in \mathcal{Q}_{\mathcal{T}}} \max_{m\in [1:M]}\Prob_{Q_{\mathcal{T}}|{\theta}}(\pmb{q}_{\mathcal{T}}|m) \biggl)\nonumber\\
        &\overset{}{=} \log \biggl(\sum_{k=1}^{M} \sum_{\substack{\pmb{q}_{\mathcal{T}} \in \mathcal{Q}_{\mathcal{T}}:\nonumber\\ \#_k \in \pmb{q}_{\mathcal{T}}}} \max_{m\in [1:M]}\Prob_{Q_{\mathcal{T}}^{\theta}|{\theta}}(\pmb{q}_{\mathcal{T}}|m)\nonumber\\
        &~~~~~~~~~+ \sum_{\substack{\pmb{q}_{\mathcal{T}} \in \mathcal{Q}_{\mathcal{T}}:\nonumber\\ \pmb{q}_{\mathcal{T}}=(\phi,\phi,\hdots,\phi)}} \max_{m\in [1:M]}\Prob_{Q_{\mathcal{T}}^{\theta}|{\theta}}(\pmb{q}_{\mathcal{T}}|m)\nonumber\\
        &~~~~~~~~~+ \sum_{m' = 1}^{M-1} \sum_{\substack{\pmb{q}_\mathcal{T}\in \mathcal{Q}_{\mathcal{T}}: \\|{\cal M}_{\pmb{q}_\mathcal{T}}|=m'+1}}\max_{m\in [1:M]}\Prob_{Q_{\mathcal{T}}^{\theta}|{\theta}}(\pmb{q}_{\mathcal{T}}|m) \biggl).\label{eqn:proofof_maxL_TSJ:stage1}
    \end{align}
    Now, we observe the following using Claim \ref{claim:thetagivenQ_Tcoll} and the structure of the protocol.
    \begin{itemize}
        \setlength{\itemsep}{5pt}
        \item The number of $T$-tuples of queries such that $\#_k\in \pmb{q}_{\mathcal{T}}$ is precisely $T$ (as $\#_k$ can appear at each of the $T$ positions with $\phi$ in the remaining $T-1$ positions). In this case, we have $
        \max_{m\in[1:M]} \Pr(\theta=m|Q_{\mathcal{T}}^{\theta}=\pmb{q}_{\mathcal{T}}) = \Pr(\theta=k|Q_{\mathcal{T}}^{\theta}=\pmb{q}_{\mathcal{T}}) = 1$. Further, we have 
        \begin{align*}
        \sum_{\substack{\pmb{q}_{\mathcal{T}} \in \mathcal{Q}_{\mathcal{T}}:\nonumber\\ \#_k \in \pmb{q}_{\mathcal{T}}}}\Pr(Q_{\mathcal{T}}^{\theta}=\pmb{q}_{\mathcal{T}})
        &=\Prob_{\theta}(k)\Prob_{M'}(0)\Pr(S\in{\cal T})\\
        &=(\Prob_{M'}(0)T)/(MN).
        \end{align*}
        \item There is precisely one query $T$-tuple such that $\pmb{q}_{\mathcal{T}}=(\phi, \phi, \hdots, \phi)$. In this case, we have $\max_{m\in[1:M]}\Prob_{Q_{\mathcal{T}}^{\theta}|\theta}(m|\pmb{q}_{\mathcal{T}})=1/M$. Also, in this case, $\Pr(Q_{\mathcal{T}}^{\theta}=\pmb{q}_{\mathcal{T}})=\Prob_{M'}(0)\Pr(S\notin{\cal T})=\Prob_{M'}(0)(1-T/N)$.
        \item If $|\mathcal{M}_{\pmb{q}_{\mathcal{T}}}|$$>1$ (which happens only if $M'=|{\cal M}_{\pmb{q}_T}|-1>0$), then $\max_{[m\in[1:M]}\Prob_{\theta | Q_{\mathcal{T}}^{\theta}}(m|\pmb{q}_{\mathcal{T}}) = \frac{1}{|\mathcal{M}_{\pmb{q}_{\mathcal{T}}}|}$.\\
        \item Finally, we have$\sum\limits_{\substack{\pmb{q}_{\mathcal{T}}\in \mathcal{Q}_{\mathcal{T}}:\nonumber\\|{\cal M}_{\pmb{q}_{\mathcal{T}}}|=m'+1}}\Prob_{Q_{\mathcal{T}}^{\theta}}(\pmb{q}_{\mathcal{T}})=\Prob_{M'}(m').$
    \end{itemize}
    Using these in \eqref{eqn:proofof_maxL_TSJ:stage1} along with the fact that $\Pr_{Q_{\cal T}^{\theta}|\theta}(\pmb{q}_{\cal T}|m)=\Prob_{\theta|Q_{\cal T}^{\theta}}(m|\pmb{q}_{\cal T})\Prob_{Q_{\cal T}^{\theta}}(\pmb{q}_{\cal T})/\Prob_{\theta}(m)=M\Prob_{\theta|Q_{\cal T}^{\theta}}(m|\pmb{q}_{\cal T})\Prob_{Q_{\cal T}^{\theta}}(\pmb{q}_{\cal T})$, we have 
    \begin{align}
        \mathsf{MaxL}(\theta,&Q_{\mathcal{T}}^{\theta})\nonumber\\
        &\overset{}{=} \log \biggl(\sum_{k = 1}^{M}\frac{\Prob_{M'}(0)T}{N} + \Prob_{M'}(0)\biggl(1-\frac{T}{N}\biggl) \nonumber\\
        &~~~~~~~~+ \sum_{m'=1}^{M-1} \frac{M}{m'+1}\Prob_{M'}(m') \biggl)\nonumber\\
        &\overset{}{=} \log M + \log\biggl(\expec\left[\frac{1}{M'+1}\right] \nonumber\\ 
        \nonumber
        &~~~~~~~~- \left(1-\frac{1}{M}\right)\left(1- \frac{T}{N}\right)P_{M'}(0) \biggl),
    \end{align}
    which completes the proof.
\end{IEEEproof}
\begin{remark}
We remark that the expressions for the rate \eqref{eqn:rate_Tcollusion_WPIR} and the privacy metrics in Lemmas \ref{lemma:TWSJ_MI_leakage} and \ref{lemma:TWSJ_Max_leakage} are identical to the expressions for the corresponding quantities in the MDS-PIR, collusion-free scenario (with $T$ in the place of $K$). 
\end{remark}

Using Lemmas \ref{lemma:TWSJ_MI_leakage} and \ref{lemma:TWSJ_Max_leakage} with specific distributions for $M'$, we now obtain the rate-privacy trade-offs for the $T$-collusion scenario in Theorems \ref{thm:TWSJ_MIL} and \ref{thm:TWSJ_MaxL}. To prove Theorem \ref{thm:TWSJ_MIL}, we choose the distribution $\Prob_{M'}(0)=\min(\frac{\rho N}{T\log M},1)=1-\Prob_{M'}(M-1)$  For Theorem \ref{thm:WBU_MaxL},  we choose $\Prob_{M'}(0)=\min(\frac{(2^{\rho}-1) N}{T(M-1)},1)=1-\Prob_{M'}(M-1).$
We omit the rest of the proofs, as they are almost identical to those in Section \ref{sec:sjwpircollusionfree}.  
\begin{theorem}
    \label{thm:TWSJ_MIL}
    For any $\rho\geq 0$, the protocol $\wcsj$ achieves the following rate and MI leakage $(\mileak, R)$ pairs.
    \begin{itemize}
        \item $\mileak = \min (\rho, \frac{T\log M}{N})$,
        \item $R = \left(1+ \left(1-\frac{N\rho }{T\log M }\right)_{+} \left(\frac{T}{N} + \cdots + \left(\frac{T}{N}\right)^{M-1} \right)\right)^{-1}$.
    \end{itemize}
\end{theorem}
\begin{theorem}
    \label{thm:TWSJ_MaxL}
    For any $\rho\geq 0$, the protocol $\wcsj$ $\wcsj$ achieves the following rate and MaxL leakage $(\maxleak, R)$ pairs.
    \begin{itemize}
        \item $\maxleak = \min (\rho, \log (1 + \frac{T(M-1)}{N}))$,
        \item $R = \left(1+ \left(1-\frac{N(2^\rho - 1)}{T(M-1)}\right)_{+} \left(\frac{T}{N} + \cdots + \left(\frac{T}{N}\right)^{M-1} \right)\right)^{-1}$.
    \end{itemize}
\end{theorem}

\begin{figure}[htbp]
    \includegraphics[width=0.5\textwidth]{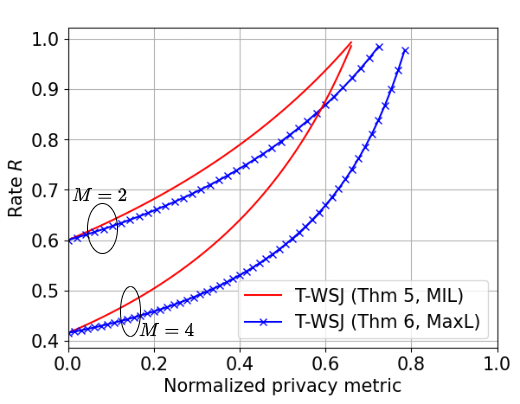}
    \caption{Trade-off between rate and privacy metrics (normalized by $\log M$) achieved by the $\wcsj$ protocol, for $N=3, T=2$ with $M\in\{2,4\}$. (Theorems \ref{thm:TWSJ_MIL} and \ref{thm:TWSJ_MaxL}).}
    \label{fig:T_WSJ}
\end{figure}

In Fig. \ref{fig:T_WSJ}, we show the rate-privacy trade-off curves for the parameters $N=3, T=2$, for $M=2$ and $M=4$. 

\section{Conclusion}
We have proposed new WPIR schemes based on the Sun-Jafar and Banawan-Ulukus schemes for the scenarios of non-colluding servers, colluding servers, and MDS-coded servers without collusion. Our WPIR schemes essentially time-share these schemes along with a clean-download phase. This simple trick achieves rate-privacy trade-offs that are either close to, or match, or even improve upon, the best known trade-offs in literature for the non-colluding and MDS-coded setups. For the server-collusion scenario, our scheme is the first WPIR scheme in literature. 

We remark on a few limitations and some ongoing work. At the present juncture, our achievable trade-offs were obtained via numerical observations. Solving for the optimal distribution of the random variable $M'$ for achieving the best possible trade-offs is an ongoing effort. On account of being lifts of the schemes in \cite{sun2017capacity,banawan2018capacity,sun2017capacityb}, our WPIR schemes inherit the issue of high file size requirement. Nevertheless, we believe that the near-universality of these schemes and their capacity-achieving nature in many known private retrieval scenarios make them worthy of consideration for future work. 
\label{sec:conclusion}

\bibliographystyle{IEEEtran}
\bibliography{WPIR_Fullversion_ArXiv.bib}
\end{document}